# FUNCTIONAL INTEGRALS FOR CORRELATED ELECTRONS

H.J. Schulz

Laboratoire de Physique des Solides
Université Paris-Sud
91405 Orsay, France

## INTRODUCTION

Functional integral methods are one way of discussing the physics of interacting fermions which in many cases turns out to be particularly transparent and appealing. In particular in cases with a broken symmetry the use of the Hubbard–Stratonovich transformation[1] often allows one to formulate the problem in a way that is both physically transparent and systematic. I here discuss some applications to the Hubbard model, both attractive and repulsive.

The standard Hubbard Hamiltonian has the form

$$H = -t \sum_{\langle \boldsymbol{r}\boldsymbol{r}' \rangle} (a^\dagger_{\boldsymbol{r}\sigma} a_{\boldsymbol{r}'\sigma} + h.c.) + U \sum_{\boldsymbol{r}} n_{\boldsymbol{r}\uparrow} n_{\boldsymbol{r}\downarrow} \quad , \tag{1}$$

where $a^\dagger_{\boldsymbol{r}\sigma}$ creates a fermion at site $\boldsymbol{r}$ with spin projection $\sigma$, $t$ is the nearest neighbor hopping integral, $U$ is the onsite interaction (either attractive or repulsive) , and $\langle \boldsymbol{r}\boldsymbol{r}' \rangle$ indicates summation over nearest–neighbor bonds, each bond being counted once. Introducing a spinor notation via

$$\Psi_{\boldsymbol{r}} = \begin{pmatrix} \psi_{\boldsymbol{r}\uparrow} \\ \psi_{\boldsymbol{r}\downarrow} \end{pmatrix} \tag{2}$$

the partition function can be written as a functional integral over Grassmann variables:

$$Z = \int \mathcal{D}\Delta e^{-S_0 - S_{int}} \quad . \tag{3}$$

The free and interaction parts of the action are respectively

$$S_0 = \int_0^\beta d\tau \left\{ \sum_{\boldsymbol{r}} \Psi^*_{\boldsymbol{r}} (\partial_\tau - \mu) \Psi_{\boldsymbol{r}} - t \sum_{\langle \boldsymbol{r}\boldsymbol{r}' \rangle} (\Psi^*_{\boldsymbol{r}} \Psi_{\boldsymbol{r}'} + c.c.) \right\} \quad , \tag{4}$$

and

$$S_{int} = U \int_0^\beta d\tau \sum_{\boldsymbol{r}} \psi^*_{\boldsymbol{r}\uparrow}\psi^*_{\boldsymbol{r}\downarrow}\psi_{\boldsymbol{r}\downarrow}\psi_{\boldsymbol{r}\uparrow} \quad . \tag{5}$$

In the following, the interaction term will be treated using a Hubbard–Stratonovich decomposition, which however is defined slightly differently for the attractive and repulsive cases.

## ATTRACTIVE HUBBARD MODEL

In this case we use the identity

$$e^{-\varepsilon U \psi^*_\uparrow \psi^*_\downarrow \psi_\downarrow \psi_\uparrow} = \frac{\varepsilon}{\pi |U|} \int d^2\Delta \exp\left[\varepsilon\left(\frac{1}{U}|\Delta|^2 - \Delta\psi^*_\uparrow\psi^*_\downarrow - \Delta^*\psi_\downarrow\psi_\uparrow\right)\right] \quad , \tag{6}$$

where $\Delta$ is a complex variable, and $\varepsilon$ is the "thickness" of the time slices used to define the functional integral.[2] Inserting this at each point in space and at each of the time slices the quartic term $S_{int}$ now has been decomposed into a purely bilinear form, but of couse there is now a functional integration over $\Delta(\boldsymbol{r},\tau)$. In the saddle point approximation for $\Delta$ one recovers BCS theory, and in the following I will derive the effective action for the low–energy excitations around the BCS ground state. For the moment I will consider a "neutral superconductor" (i.e. a superfluid) and therefore neglect the coupling to the electromagnetic filed. The standard way to do the expansion is to write

$$\Delta(\boldsymbol{r},\tau) = \Delta_0 + \delta(\boldsymbol{r},\tau) \quad . \tag{7}$$

where $\Delta_0$ is the BCS gap parameter, and then to expand in $\delta$. However, in this way amplitude and phase excitations are mixed up and some effort is needed to disentangle both and to obtain explicitly the important phase excitations. The derivation can be considerably simplified by performing first a gauge transformation on the fermions: one writes

$$\Delta(\boldsymbol{r},\tau) = |\Delta(\boldsymbol{r},\tau)| e^{i\varphi(\boldsymbol{r},\tau)} \quad , \tag{8}$$

and defines new fermion fields $\Phi$ by

$$\Psi_{\boldsymbol{r}} = e^{i\varphi(\boldsymbol{r},\tau)/2}\Phi_{\boldsymbol{r}} \quad . \tag{9}$$

The action then takes the form

$$S = S_{BCS}(\Phi,\Delta_0) + S_\varphi(\Phi,\varphi) + S_\delta(\Phi,|\Delta| - \Delta_0) \quad . \tag{10}$$

Here $S_{BCS}$ is the mean–field action, $S_\delta$ contains the amplitude fluctuations and is of minor importance as long as one is well below $T_c$ and in the weak–coupling limit (so that the coherence length is large). The important term is the contribution from the phase degrees of freedom:

$$S_\varphi = \int_0^\beta d\tau \left\{ \frac{i}{2}\sum_{\boldsymbol{r}} \dot\varphi_{\boldsymbol{r}} \Phi^*_{\boldsymbol{r}}\Phi_{\boldsymbol{r}} - t \sum_{\langle \boldsymbol{r},\boldsymbol{r}'\rangle} \left[ (e^{-i(\varphi_{\boldsymbol{r}} - \varphi_{\boldsymbol{r}'})/2} - 1)\Phi^*_{\boldsymbol{r}}\Phi_{\boldsymbol{r}'} + c.c. \right] \right\} \quad . \tag{11}$$

This is now explicitly of at least first order in the derivatives of $\varphi$, and consequently in order to obtain the lowest order effective action, one has to expand the action to second order in $S_\varphi$. The result is

$$S_{\text{eff}}(\varphi) = \frac{1}{2}\int_0^\beta d\tau \int d^d r \{\kappa\dot\varphi^2 + \rho_s(\nabla\varphi)^2\} \quad . \tag{12}$$

The coefficient $\kappa$ is related to the compressibility of the fermions in the BCS state and given by

$$\kappa = \tfrac{1}{4}\langle nn\rangle_{q,\omega=0} \quad , \tag{13}$$

and the superfluid density is

$$\rho_s = E_{kin} + \tfrac{1}{4}\langle jj\rangle_{q,\omega=0} \quad . \tag{14}$$

The expectation values are taken with respect to the BCS ground state. Note that the result for $\rho_s$ contains a contribution ($E_{kin}$, the kinetic energy per bond) which comes from a first order expectation value of $S_\varphi$. The effective action (12) obviously leads to the well–known Anderson–Goldstone collective mode[3] for the phase excitations, with energy $\omega(q) = v|q|$, $v^2 = \rho_s/\kappa$.

One of the advantages of the present formulation is that the inclusion of the external electromagnetic field is straightforeward. The scalar potential leads to an extra term

$$-eV(\boldsymbol{r},\tau)\Phi^*_{\boldsymbol{r}}\Phi_{\boldsymbol{r}} \tag{15}$$

in the fermion action, which corresponds to the replacement

$$\dot\varphi_r \rightarrow \dot\varphi_r + 2ieV(\boldsymbol{r},\tau) \tag{16}$$

in $S_\varphi$. On the other hand, the vector potential is included via the Peierls substitution

$$\Phi^*_{\boldsymbol{r}}\Phi_{\boldsymbol{r}'} \rightarrow \Phi^*_{\boldsymbol{r}}\Phi_{\boldsymbol{r}'} \exp\left(-\frac{ie}{c}\int_{\boldsymbol{r}}^{\boldsymbol{r}'} \boldsymbol{A}\cdot d\boldsymbol{\ell}\right) \quad , \tag{17}$$

which for slowly varying $\boldsymbol{A}$, i.e. small fields, gives rise to the replacement

$$\nabla\varphi \rightarrow \nabla\varphi - \frac{2e}{c}\boldsymbol{A} \quad , \tag{18}$$

and therefore the effective action is now

$$S_{\text{eff}}(\varphi) = \frac{1}{2}\int_0^\beta d\tau \int d^dr\{\kappa(\dot\varphi + 2ieV(\boldsymbol{r},\tau))^2 + \rho_s(\nabla\varphi - \frac{2e}{c}\boldsymbol{A}(\boldsymbol{r},\tau))^2\} + S_{\text{e.m.}} \quad . \tag{19}$$

Here $S_{\text{e.m.}}$ is the action of the free electromagnetic field. One may notice that this derivation is entirely free of the usual "eikonal approximation" used to introduce the electromagnetic field. The essential point now is that (19) gives, via the Anderson–Higgs mechanism,[3] rise to a shift of the energy of the long–wavelength phase oscillations away from zero to the plasmon frequency.

A number of comments are in order here. First, the derivation is limited to long wavelength, so that $q\xi \ll 1$. Otherwise the expectation values in (13) and (14) would have to be taken at nonzero $q$ and $\omega$. Further, the derivation is limited to zero temperature, because at any finite temperature excited quasiparticles give rise to contributions to $\langle nn\rangle_{q,\omega}$ which are nonanalytic as $q, \omega \rightarrow 0$. Finally, we notice that there are of course terms of higher order in $\varphi$. Given that $\nabla\varphi$ is proportional to the supercurrent, time–reversal invariance requires there to be only even powers in gradients of $\varphi$. On the other hand there is no analogous restriction for $\dot\varphi$, which represents variations of the local particle density. In particular a term proportional to $\dot\varphi(\nabla\varphi)^2$ is allowed by symmetry and certainly non–zero, because it represents the change of the kinetic energy of a supercurrent when the particle density changes. This term is in fact important to obtain the London acceleration equation, as first pointed out by Abrahams and Tsuneto.[4]

## REPULSIVE HUBBARD MODEL

For repulsive interactions ($U > 0$) the Hubbard–Stratonovich decomposition in the form (6) is obviously unsuitable because the integrals are divergent. In this case the appropriate form is

$$e^{-\varepsilon U \psi_\uparrow^* \psi_\downarrow^* \psi_\downarrow \psi_\uparrow} = \frac{\varepsilon}{\pi U} \int d\Delta_c d\Delta_s \exp\left[ -\frac{\varepsilon}{U}(\Delta_c^2 + \Delta_s^2) + i\varepsilon \Delta_c n + \varepsilon \Delta_s \sigma_z \right] \quad . \tag{20}$$

Here $\Delta_{c,s}$ are real variables, and $n = \psi_\uparrow^* \psi_\uparrow + \psi_\downarrow^* \psi_\downarrow$, $\sigma_z = \psi_\uparrow^* \psi_\uparrow - \psi_\downarrow^* \psi_\downarrow$. As in the attractive case, one inserts this at each point in space and time and thus obtains a functional integral over charge and spin fields $\Delta_{c,s}(r, \tau)$, coupled bilinearly to the fermions. A saddle point approximation reproduces the Hartree–Fock results, and in particular at half–filling one finds an antiferromagnetic (or spin–density wave in another terminology) ground state. The unpleasant feature of this way of proceeding is that both $\Delta_c$ and $\Delta_s$ are scalar fields, and one therefore cannot construct easily the effective action for the low–energy excitations of the antiferromagnetic state which are spin–waves, the existence of which is of course closely related to the vectorial character of the order parameter.

Alternatively, one might use a Hubbard–Stratonovich decomposition using a vector auxiliary field. One then however does not even obtain the Hartree–Fock solution as a saddle point. A number of other, equally unsatisfactory decompositions have been discussed in the literature.[5] In order to obtain a spin–rotation invariant effective action for fluctuations around the Hartree–Fock solution, I notice[6,7] that in writing down the Hamiltonian the choice of the spin quantization axis is a priori arbitrary at each lattice site, and in a functional integral formulation can also vary in time. I then leave the quantization axis $\mathbf{\Omega}_{\boldsymbol{r}}(\tau)$ arbitrary and integrate over all possible $\mathbf{\Omega}_{\boldsymbol{r}}(\tau)$, with the appropriate invariant and normalized integration measure at each point in space and time. In practice, this is achieved by introducing $SU(2)$ rotation matrices $R_{\boldsymbol{r}}(\tau)$ at each point of space and time which satisfy $R_{\boldsymbol{r}}(\tau)\sigma_z R_{\boldsymbol{r}}^+(\tau) = \mathbf{\Omega}_{\boldsymbol{r}}(\tau) \cdot \boldsymbol{\sigma}$. A convenient choice is

$$R(\mathbf{\Omega}) = \begin{pmatrix} \cos(\frac{1}{2}\theta) & -e^{-i\varphi}\sin(\frac{1}{2}\theta) \\ e^{i\varphi}\sin(\frac{1}{2}\theta) & \cos(\frac{1}{2}\theta(\boldsymbol{r}, \tau)) \end{pmatrix} \quad , \tag{21}$$

where $\theta$ and $\varphi$ are the usual polar angles. I then introduce identities $R_{\boldsymbol{r}}(\tau)R_{\boldsymbol{r}}^+(\tau) = 1$ at the appropriate places in the functional integral and integrate over all configurations $\mathbf{\Omega}_{\boldsymbol{r}}(\tau)$. Finally, new spinor varibles are introduced via

$$\Phi = R^+ \Psi \quad . \tag{22}$$

This means that the $\phi$–particles now have their spin along $\pm\mathbf{\Omega}_{\boldsymbol{r}}(\tau)$. The Hubbard interaction term is invariant under this transformation, and now the Hubbard–Stratonovich transformation can be used in its form (20) without loosing the spin excitations which are contained in the functional integral over $\mathbf{\Omega}_{\boldsymbol{r}}(\tau)$. This also means that a nonzero saddle point value of the spin field $\Delta_s$ does not necessarily imply the existence of magnetic long–range order. For this to occur, the angular degrees of freedom have also to be ordered.

### Half–Filled Case

At half–filling, the Hartree-Fock saddle point is antiferromagnetic for any positive $U$. The partition function then becomes

$$Z = \int \mathcal{D}\mathbf{\Omega}\, \mathcal{D}\Phi\, \mathcal{D}\delta\, e^{-S_{HF}-S_\Omega-S_\delta} \quad . \tag{23}$$

Here $S_{HF}$ is the action corresponding to the saddle point, $S_\delta$ represents the massive fluctuations of $\Delta_c$ and $\Delta_s$ around their respective saddle point values, and $S_\Omega$ represents the coupling between the angular fluctuations and the fermions and is given by

$$S_\Omega = \int_0^\beta d\tau \{\Phi^*_{\boldsymbol{r}} R^+_{\boldsymbol{r}} \dot{R}_{\boldsymbol{r}} \Phi_{\boldsymbol{r}} - t \sum_{\langle \boldsymbol{r}\boldsymbol{r}'\rangle} [\Phi^*_{\boldsymbol{r}} (R^+_{\boldsymbol{r}} R_{\boldsymbol{r}'} - 1)\Phi_{\boldsymbol{r}'} + c.c.]\} \quad . \tag{24}$$

This term is explicitly of at least first order in time and space derivatives. An effective action for $\mathbf{\Omega}_{\boldsymbol{r}}(\tau)$ can be obtained by integrating out the fermionic degrees of freedom, and to second order one finds

$$S_{\text{eff}}(\mathbf{\Omega}) = \langle S_\Omega \rangle_{HF} - \tfrac{1}{2}\langle S_\Omega^2 \rangle_{HF,con} \quad . \tag{25}$$

Note that there are explicitly nonzero first order cumulants. Dividing $S_\Omega$ into two parts $S_{\Omega,\tau}$ and $S_{\Omega,\boldsymbol{r}}$ containig only time or space derivatives, respectively, one finds in particular

$$\begin{aligned} \langle S_{\Omega,\tau}\rangle_{HF} &= \frac{i}{2}\int_0^\beta d\tau\ m_0 \sum_{\boldsymbol{r}} (-1)^{\boldsymbol{r}} (1-\cos\theta_{\boldsymbol{r}})\dot{\varphi}_{\boldsymbol{r}} \\ &= \frac{i m_0}{2} \sum_{\boldsymbol{r}} (-1)^{\boldsymbol{r}} \int_0^\beta \boldsymbol{A}(\mathbf{\Omega}) \cdot d\mathbf{\Omega} \quad . \end{aligned} \tag{26}$$

Here $m_0$ is the SDW amplitude which tends to unity for large $U$, $\theta_{\boldsymbol{r}}$ and $\varphi_{\boldsymbol{r}}$ are the polar angles on site $\boldsymbol{r}$, and $\boldsymbol{A}(\mathbf{\Omega})$ is the vector potential created by a magnetic monopole sitting at the center of a unit sphere.[8,9] Eq.(26) can be recognized as a collection of the Berry phase terms for spins localized at lattice sites $\boldsymbol{r}$. The standard procedure[8,9] is now to assume at least local antiferromagnetic order and to write

$$\mathbf{\Omega}_{\boldsymbol{r}} = \sqrt{1 - a^2 \boldsymbol{L}^2_{\boldsymbol{r}}}\boldsymbol{n}_{\boldsymbol{r}} + (-1)^{\boldsymbol{r}} a \boldsymbol{L}_{\boldsymbol{r}} \ , \quad \boldsymbol{n}_{\boldsymbol{r}} \cdot \boldsymbol{L}_{\boldsymbol{r}} = 0 \quad . \tag{27}$$

Here the antiferromagnetic order parameter field $\boldsymbol{n}_{\boldsymbol{r}}$ is slowly varying in space and time, and $\boldsymbol{L}_{\boldsymbol{r}}$ which represents magnetic fluctuations around $\boldsymbol{q} = 0$ is small. $a$ is the lattice constant. Inserting this into eq.(26) one obtains (I here specialize to the two–dimensional case)

$$\langle S_{\Omega,\tau}\rangle_{HF} = \frac{i m_0}{2a} \int d^2 r\, d\tau \boldsymbol{L} \cdot (\boldsymbol{n} \times \dot{\boldsymbol{n}}) \quad . \tag{28}$$

For large $U$ this term dominates the spin dynamics, however, for moderate or small $U$ the $\langle S^2_{\Omega,\tau}\rangle_{HF}$ term becomes more important. Expanding now all terms up to second order, the effective action is obtained as

$$S_{\text{eff}} = \frac{1}{2}\int_0^\beta d\tau \int d^2 r \{\alpha_1 \boldsymbol{L}^2 + i\alpha_2 \boldsymbol{L}\cdot(\boldsymbol{n}\times\dot{\boldsymbol{n}}) + \alpha_3 \dot{\boldsymbol{n}}^2 + \rho_s (\nabla \boldsymbol{n})^2\} \quad . \tag{29}$$

The correlation strength enters via the $U$–dependence of the coefficients. In particular one has

$$\rho_s = E_{kin} + \tfrac{1}{4}\langle j^+ j^- \rangle_{\boldsymbol{q}=\omega=0} \quad , \tag{30}$$

where $j^\pm$ are the transverse spin current operators. The similarity with eq.(14) for the superconducting case is obvious. Using the Hartree–Fock equations of motion, one can show that current–current correlaion functions are closely related to the spin–spin correlation functions which appear in previous results.[10–12] Using the notations of Schrieffer *et al.*[10] and of Chubukov and Frenkel,[12] the coefficients are given by

$$\alpha_1 = 2\Delta^2\left[\frac{1}{U} - \chi_0^{+-}(0,0)\right] = 2\Delta^2 z \tag{31}$$

$$\alpha_2 = \frac{4\Delta^2}{a}\left.\frac{\partial}{\partial\omega}\chi_Q^{+-}(0,\omega)\right|_{\omega=0} = \frac{2\Delta^3 x}{a} \tag{32}$$

$$\alpha_3 = \frac{\Delta^2}{a^2}\left.\frac{\partial^2}{\partial\omega^2}\chi_0^{+-}(\boldsymbol{Q},\omega)\right|_{\omega=0} = \frac{\Delta^2 x}{2a^2} \tag{33}$$

$$\rho_s = -\frac{\Delta^2}{a^2}\left.\frac{\partial^2}{\partial q_x^2}\chi_0^{+-}(\boldsymbol{q},0)\right|_{\boldsymbol{q}=\boldsymbol{Q}} = 2t^2\Delta^2 y \quad . \tag{34}$$

Here $\Delta$ is the SDW gap parameter, and

$$x = \frac{1}{N}\sum_{\boldsymbol{k}}{}' \frac{1}{E_{\boldsymbol{k}}^3} = \frac{1}{2\Delta^3}\, {}_3F_2(\tfrac{1}{2},\tfrac{1}{2},\tfrac{3}{2};1,1;\tau) \quad , \tag{35}$$

$$y = \frac{1}{N}\sum_{\boldsymbol{k}}{}' \frac{\sin^2(k_x a)}{E_{\boldsymbol{k}}^3} = \frac{1}{4\Delta^3}\, {}_3F_2(\tfrac{1}{2},\tfrac{3}{2},\tfrac{3}{2};2,2;\tau) \quad , \tag{36}$$

$$z = \frac{1}{N}\sum_{\boldsymbol{k}}{}' \frac{\varepsilon_{\boldsymbol{k}}^2}{E_{\boldsymbol{k}}^3} = \frac{2t^2}{\Delta^3}\, {}_3F_2(\tfrac{3}{2},\tfrac{3}{2},\tfrac{3}{2};2,2;\tau) \quad . \tag{37}$$

The ${}_3F_2$ are generalized hypergeometric functions, and $\tau = -16t^2/\Delta^2$.

Integrating now over $\boldsymbol{L}$ in (29) one obtains the well–known action of the nonlinear sigma model

$$S_{\text{eff}} = \frac{1}{2}\int_0^\beta d\tau \int d^2r \{\alpha\dot{\boldsymbol{n}}^2 + \rho_s(\nabla\boldsymbol{n})^2\} \quad , \tag{38}$$

with $\alpha = \alpha_2^2/(4\alpha_1) + \alpha_3$. From this form of the effective action we immediately obtain an expression for the spin–wave velocity:

$$c^2 = \frac{4t^2a^2yz}{\Delta^2x^2 + xz} \quad . \tag{39}$$

Numerical results for the spin–wave velocity and the spin stiffness as a function of $U$ are shown in the figures. In obtaining these results it proved useful (especially for small $U$) to transform the hypergeometric functions in eqs.(35) to (37) into complete elliptic integrals, using formulae given by Prudnikov *et al.*.[13]

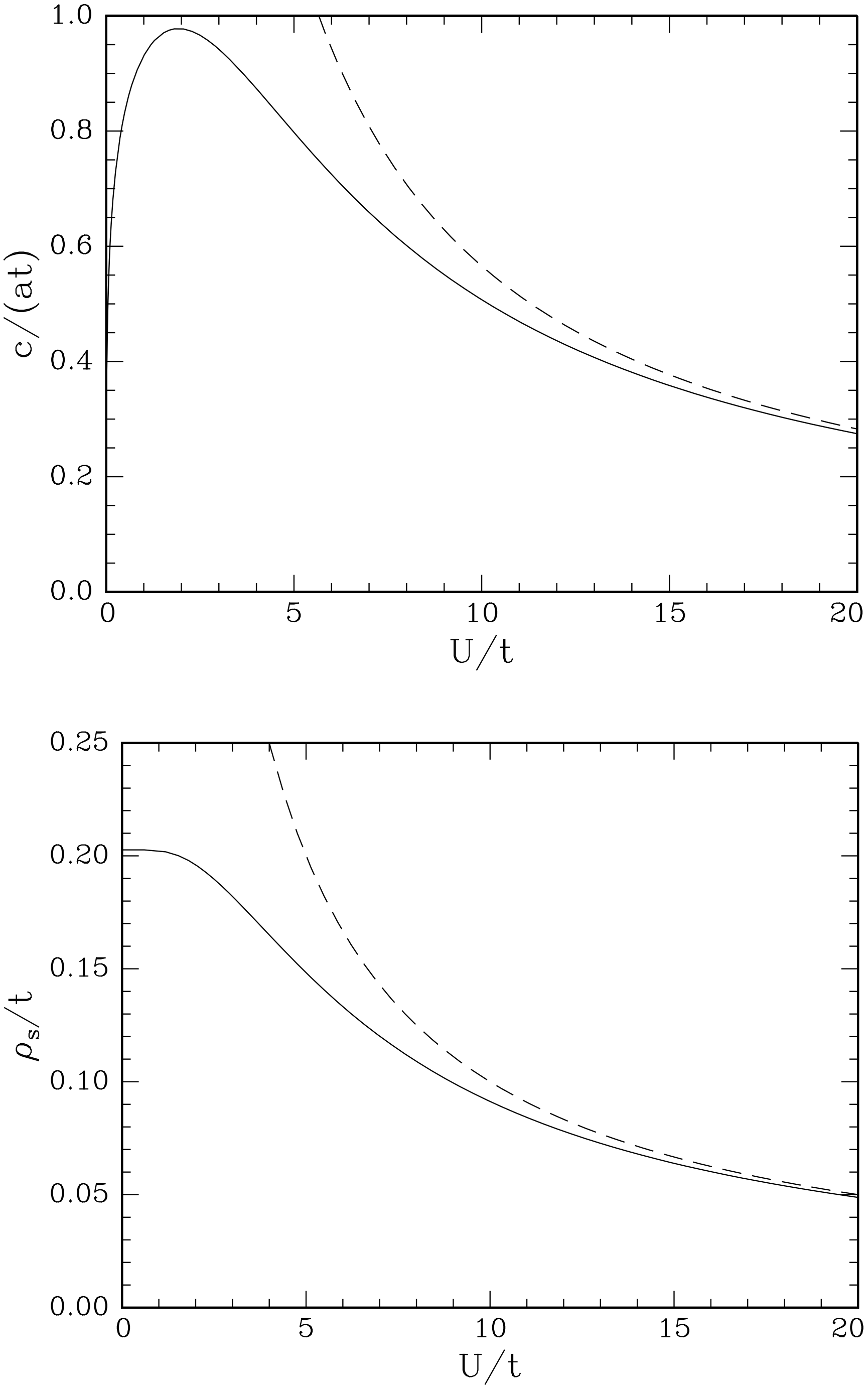

FIG. 1. Spin–wave velocity (top) and spin stiffness (bottom) for the two–dimensional Hubbard model at half–filling as a function of $U$. The dashed lines are the asymptotic forms for large $U$, $c = 4\sqrt{2}t^2a/U$ and $\rho_s = t^2/U$, respectively, which agree with lowest order results for the antiferromagnetic Heisenberg model.

One should note that the correlation strength also enters via the value of the short–distance (i.e. large–$q$) cutoff beyond which antiferromagnetic short–range order can not be defined. For large $U$ this cutoff is the lattice constant, however, for general $U$ this cutoff can be identified with the coherence length $\xi_0 = t/\Delta$ (the analogue of the BCS coherence length of superconductivity). In particular, this length becomes large for small $U$ and diverges as

$\exp(\sqrt{t/U})$ for $U \to 0$. Consequently the coupling constant of the nonlinear sigma model[14]

$$g = c/(\rho_s \xi_0) \tag{40}$$

goes to zero as $U \to 0$, i.e. *quantum spin fluctuation corrections to the Hartree–Fock solution become arbitrarily small for small $U$.* The same argument also applies to the amplitude fluctuations contained in $S_\delta$. In particular, the small–$U$ limit of $\rho_s$ is thus expected to be an exact result. On the other hand, for larger $U$ there are corrections both to $c$ and to $\rho_s$, as is well–known from the large–$U$ limit which is equivalent to the Heisenberg model.[14] In this limit the above derivation can actually be formulated so that arbitrary (and not only slow long–wavelength) variations of $\mathbf{\Omega}_{\boldsymbol{r}}$ in space and time are allowed, and one then directly recovers the action of the antiferromagnetic spin–1/2 Heisenberg model.[6]

In two dimensions, thermally excited spin–wave fluctuations destroy antiferromagnetic long–range order at any nonzero temperature. At low temperatures the dominant fluctuations are always the orientational spin excitations, described by the effective action (38) (with possibly weakly temperature dependent coefficients). It is interesting to note that for small $U$ there is a very abrupt crossover at the mean–field transition temperature $T_c^{MF}$ between this regime and a high–temperature Fermi liquid state. The width of the crossover is of order $(T_c^{MF})^2/t$.[15] For larger $U$ this crossover gets more and more smeared out.

**The Doped Case**

Doping away from half–filling, the antiferromagnetic ground state of the Hubbard model is modified quite drastically: for small $U$ the mean–field (saddle point) approach predicts that for any finite doping the Néel state is replaced by a linearly polarized incommensurate spin–density wave.[16,17] For small doping the magnetic structure is best described as a regular arrangement of linear domain walls separating commensurate regions with opposite signs of the order parameter. In this case the doped carriers are localized at the domain walls, and the systems thus remains insulating. With increasing doping, the domain wall state progressively transforms into a sinusoidal modulation, and this is accompanied by a metal–insulator transition.[16] It would clearly be interesting to generalize the above treatment of the half–filled case to investigate the collective modes in the incommensurate state.

The mean–field picture has been critized by Chubukov and Frenkel,[12] who argue that corrections to mean–field suppress the instability against domain wall formation at least at small doping, and that therefore the commensurate antiferromagnetic state survives even at finite doping. However, their argument is entirely based on a *local* stability analysis of the antiferromagnetic state, and this may well be insufficient when one considers domain wall formation which implies a *global* reorganization of the magnetization pattern. In fact, within mean–field the energy gain per doped particle due to domain wall formation is of order $\Delta$,[16] and for small $U$ corrections to mean–field are typically of order $\Delta^2/t$. This strongly suggests that the mean–field picture is correct for weak correlation.

For large $U$ an effective action for the spin degrees of freedom and the doped carriers can be derived, because in fact arbitrary space–time variations of $\mathbf{\Omega}(\boldsymbol{r}, \tau)$ can be treated. For simplicity, one can then start from a ferromagnetic saddle point which is characterized by lower and upper Hubbard band separated by a gap $U$. For the case of electron doping, the chemical potential is somewhere in the upper Hubbard bands, and the lower Hubbard band then can be integrated out.[6] In this way one obtains the effective action for the local spin orientation and particles in the upper Hubbard band order by order in $t/U$. To zeroth order

in $t/U$ I find

$$S_{eff}^0 = \int_0^\beta d\tau \left\{ \sum_{\boldsymbol{r}} [\phi_{\boldsymbol{r}}^* (\partial_\tau - \mu + U)\phi_{\boldsymbol{r}} - \frac{i}{2}\dot{\varphi}_{\boldsymbol{r}}(1-\cos\vartheta_{\boldsymbol{r}})(1-\phi_{\boldsymbol{r}}^*\phi_{\boldsymbol{r}})] \right.$$
$$\left. -t \sum_{\langle \boldsymbol{r}\boldsymbol{r}'\rangle} [\alpha(\boldsymbol{\Omega}_{\boldsymbol{r}}, \boldsymbol{\Omega}_{\boldsymbol{r}'})\phi_{\boldsymbol{r}}^*\phi_{\boldsymbol{r}'} + c.c.] \right\} \quad . \tag{41}$$

Here $\phi$ refers to fermions in the upper Hubbard band, the spin index being omitted, and $\varphi_{\boldsymbol{r}}, \vartheta_{\boldsymbol{r}}$ are the polar angles of $\boldsymbol{\Omega}_{\boldsymbol{r}}$. The coefficients $\alpha(\boldsymbol{\Omega}_{\boldsymbol{r}}, \boldsymbol{\Omega}_{\boldsymbol{r}'})$ come from the expression for the product of two $R$ matrices:

$$R(\boldsymbol{\Omega})R^+(\boldsymbol{\Omega}') = \begin{pmatrix} \alpha(\boldsymbol{\Omega},\boldsymbol{\Omega}') & e^{-i\varphi'}\alpha(\boldsymbol{\Omega},-\boldsymbol{\Omega}') \\ -e^{i\varphi'}\alpha^*(\boldsymbol{\Omega},-\boldsymbol{\Omega}') & \alpha^*(\boldsymbol{\Omega},\boldsymbol{\Omega}') \end{pmatrix} \quad , \tag{42}$$

where

$$\alpha(\boldsymbol{\Omega}_{\boldsymbol{r}}, \boldsymbol{\Omega}_{\boldsymbol{r}'}) = |\alpha| e^{i\chi_{\boldsymbol{r}\boldsymbol{r}'}} = [(1+\boldsymbol{\Omega}_{\boldsymbol{r}}\cdot\boldsymbol{\Omega}_{\boldsymbol{r}'})/2]^{1/2} \exp[i\hat{A}(\boldsymbol{\Omega}_{\boldsymbol{r}}, \boldsymbol{\Omega}_{\boldsymbol{r}'}, \hat{\boldsymbol{z}})/2] \quad , \tag{43}$$

$\hat{A}(\boldsymbol{\Omega}_1, \boldsymbol{\Omega}_2, \boldsymbol{\Omega}_3)$ is the signed solid angle spanned by the vectors $\boldsymbol{\Omega}_1, \boldsymbol{\Omega}_2, \boldsymbol{\Omega}_3$[18], and $\hat{\boldsymbol{z}}$ is the unit vector along $z$.

In the absence of particles in the upper Hubbard band, in $S_{eff}^0$ only the purely imaginary term remains, which is the Berry phase of an isolated spin $1/2$, i.e., as expected, the half-filled Hubbard model becomes a collection of independent spins for $U = \infty$. Introducing more fermions, two effects occur: (i) the factors $(1 - \phi_{\boldsymbol{r}}^*\phi_{\boldsymbol{r}})$, previously introduced by Shankar from semi-phenomenological arguments[9], cancel the Berry phase term whenever there is an extra particle on site $\boldsymbol{r}$, i.e. one is in a spin singlet whenever two particles occupy the same site. Here this effect is seen directly from a microscopic calculation. (ii) the kinetic energy term plays a rôle: in particular, going around an elementary plaquette $(1234)$ the lattice curl of the phases $\chi_{\boldsymbol{r}\boldsymbol{r}'}$ equals $\Phi_{1234} = [\hat{A}(\boldsymbol{\Omega}_1, \boldsymbol{\Omega}_2, \boldsymbol{\Omega}_3) + \hat{A}(\boldsymbol{\Omega}_3, \boldsymbol{\Omega}_4, \boldsymbol{\Omega}_1)]/2$, i.e. there is an effective magnetic field proportional to the solid angle spanned by $\boldsymbol{\Omega}_1, ..., \boldsymbol{\Omega}_4$. $\Phi_{1234}$ is the lattice analogue of the familiar winding number density of the continuum nonlinear $\sigma$ model[18]. Note that, while the gauge potential in (43) depends explicitly on $\hat{\boldsymbol{z}}$ and therefore is not rotational invariant, the physical fluxes are. For coplanar configurations $\Phi_{1234} = 0$, i.e. the phases can be removed by a gauge transformation of the $\phi$'s. One then straightforwardly sees that the kinetic term is optimized by a ferromagnetic arrangement of the spins. This is the familiar Nagaoka phenomenon.[19] Whether non-coplanar configurations of $\boldsymbol{\Omega}_{\boldsymbol{r}}$ with a nonzero winding number density can lead to an energy lower than the Nagaoka state is not currently clear.[20]

The first order contribution to the action is

$$S_{eff}^1 = \frac{t^2}{U}\int_0^\beta d\tau \left\{ \frac{1}{2}\sum_{\langle \boldsymbol{r}\boldsymbol{r}'\rangle} (2 - \phi_{\boldsymbol{r}}^*\phi_{\boldsymbol{r}} - \phi_{\boldsymbol{r}'}^*\phi_{\boldsymbol{r}'})(\boldsymbol{\Omega}_{\boldsymbol{r}}\cdot\boldsymbol{\Omega}_{\boldsymbol{r}'} - 1) \right.$$
$$+ \sum_{\langle \boldsymbol{r}\boldsymbol{r}'\boldsymbol{r}''\rangle} [\alpha(\boldsymbol{\Omega}_{\boldsymbol{r}}, -\boldsymbol{\Omega}_{\boldsymbol{r}'})\alpha(-\boldsymbol{\Omega}_{\boldsymbol{r}'}, \boldsymbol{\Omega}_{\boldsymbol{r}''})\phi_{\boldsymbol{r}}^*\phi_{\boldsymbol{r}''} + c.c.]$$
$$\left. + \frac{1}{4t^2}\sum_{\boldsymbol{r}} (1 - \phi_{\boldsymbol{r}}^*\phi_{\boldsymbol{r}})\dot{\boldsymbol{\Omega}}_{\boldsymbol{r}}^2 \right\} \quad . \tag{44}$$

Here $\boldsymbol{r}$ and $\boldsymbol{r}''$ are second– or third–nearest neighbors, and the sum over $\boldsymbol{r}'$ is over all sites that are nearest neighbors of both $\boldsymbol{r}$ and $\boldsymbol{r}''$. In the absence of fermions in the upper Hubbard band, only the $\phi$–independent part of the first term in $S_{eff}^1$ contributes and represents the antiferromagnetic exchange interaction between nearest neighbor sites, e.g. in this case $S^0_{eff} + S^1_{eff}$ is the action of the antiferromagnetic Heisenberg model, as already mentioned above.

In the presence of a finite concentration $n$ of extra electrons, one immediately sees the instability of the Néel state: in the Néel state the term proportional to $t$ in $S^0_{eff}$ does not contribute. If however $\mathbf{\Omega}_{\boldsymbol{r}} \cdot \mathbf{\Omega}_{\boldsymbol{r}'} = -1 + \varepsilon^2$, there is an effective nearest neighbor hopping of order $t|\varepsilon|$, and a corresponding gain of kinetic energy of order $-t|\varepsilon|n$. The loss of exchange energy is of order $\varepsilon^2$, e.g. $\varepsilon \neq 0$ is energetically favored for any nonzero $n$. For an $\boldsymbol{r}$-independent $\varepsilon$ one then finds a spiral, as proposed by Shraiman and Siggia[21]. However, at least at the mean–field level discussed here the spiral state is unstable against phase separation.[6]

A word of caution is in place here: as describrd above, the a formalism neglects certain important contributions to the functional integral. This mainly comes from the fact that $\downarrow$ particle (lower Hubbard band) can equally well described by an $\uparrow$ particle (upper Hubbard band), provided that the local $\mathbf{\Omega}_{\boldsymbol{r}}$ is changed into $-\mathbf{\Omega}_{\boldsymbol{r}}$. Ways to handle this problem will be discussed in a forthcoming publication.

## GENERALIZATIONS

### $\boldsymbol{SU(2)} \times \boldsymbol{SU(2)}$ Symmetry

The present approach allows an interesting generalization: using the matrix representation[22]

$$\Psi_{\boldsymbol{r}} = \begin{pmatrix} \psi_{\boldsymbol{r}\uparrow} & \psi_{\boldsymbol{r}\downarrow} \\ (-1)^{\boldsymbol{r}}\psi^*_{\boldsymbol{r}\downarrow} & -(-1)^{\boldsymbol{r}}\psi^*_{\boldsymbol{r}\uparrow} \end{pmatrix} \quad . \tag{45}$$

the Hamiltonian can be written in a $SU(2) \times SU(2)$ invariant way:[6,23] multiplying $\Psi_{\boldsymbol{r}}$ by a $SU(2)$ matrix from the right, one generates the spin rotations discussed above ($SU(2)_s$ symmetry). On the other hand, multiplication from the left ($SU(2)_c$ symmetry) generates electron-hole transformations (in the Heisenberg model this becomes a $SU(2)$ gauge symmetry[22]). The clear separation between the two symmetries makes the representation (45) very useful.

The reason for introducing the factors $(-1)^{\boldsymbol{r}}$ in (45) is the identity, *valid if $\boldsymbol{r}$ and $\boldsymbol{r}'$ are on different sublattices*:

$$\mathrm{tr}[\Psi^+_{\boldsymbol{r}}\Psi_{\boldsymbol{r}'}] = \sum_s (\psi^*_{\boldsymbol{r}s}\psi_{\boldsymbol{r}'s} + \psi^*_{\boldsymbol{r}'s}\psi_{\boldsymbol{r}s}) \quad . \tag{46}$$

Using these relations, it is straightforward to rewrite the Hamiltonian in a form invariant under $SU(2)_c \times SU(2)_s$. One possible form is

$$H(\psi^*, \psi) = -t \sum_{\langle \boldsymbol{r}\boldsymbol{r}'\rangle s} \mathrm{tr}[\Psi^+_{\boldsymbol{r}}\Psi_{\boldsymbol{r}'}] - \frac{U}{24}\sum_{\boldsymbol{r}} \mathrm{tr}[\boldsymbol{\sigma}\Psi^+_{\boldsymbol{r}}\Psi_{\boldsymbol{r}}] \cdot \mathrm{tr}[\boldsymbol{\sigma}\Psi^+_{\boldsymbol{r}}\Psi_{\boldsymbol{r}}] \quad . \tag{47}$$

The action becomes

$$S = \int_0^\beta d\tau \left[\frac{1}{2}\sum_{\boldsymbol{r}} \mathrm{tr}[\Psi^+_{\boldsymbol{r}}(\partial_\tau - \mu\sigma_z)\Psi_{\boldsymbol{r}}] + H(\psi^*, \psi)\right] \quad , \tag{48}$$

One sees immediately that $\mu \neq 0$ breaks the $SU(2)_c \times SU(2)_s$ symmetry, with only $U(1)_c \times SU(2)_s$ (multiplication by a phase and spin rotation) left.

I now perform the unitary transformation $\Phi_{\boldsymbol{r}} = R_c \Psi_{\boldsymbol{r}} R_s$, where $R_c, R_s$ are $SU(2)$ matrices, parametrized by unit vectors $\boldsymbol{\Omega}_c, \boldsymbol{\Omega}_s$ which vary in time and space, similarly to eq.(21):

$$R_c^+ \sigma_z R_c = \boldsymbol{\Omega}_c \cdot \boldsymbol{\sigma} \ , \quad R_s \sigma_z R_s^+ = \boldsymbol{\Omega}_s \cdot \boldsymbol{\sigma}^T \quad . \tag{49}$$

A $\Phi$ particle with spin up points along $\boldsymbol{\Omega}_s$ in the original (laboratory) frame. Similarly, for $R_c \neq 1$ a $\phi$ creation operator is a linear combination of the original creation and destruction operators. In terms of $\Phi$ one has

$$\begin{aligned} &S(\phi^*, \phi, \boldsymbol{\Omega}_c, \boldsymbol{\Omega}_s) = \\ &\quad \int_0^\beta d\tau \left[ \frac{1}{2} \sum_{\boldsymbol{r}} \mathrm{tr}[\Phi_{\boldsymbol{r}}^+ (\partial_\tau \Phi_{\boldsymbol{r}} + R_c \, (\partial_\tau R_c^+) \Phi_{\boldsymbol{r}} + \Phi_{\boldsymbol{r}} (\partial_\tau R_s^+) R_s \, - \mu R_c \, \sigma_z R_c^+ \Phi_{\boldsymbol{r}})] \right. \\ &\qquad \left. + H(\phi^*, \phi, \boldsymbol{\Omega}_c, \boldsymbol{\Omega}_s) \right] \quad . \end{aligned} \tag{50}$$

Now, introducing at each point in space and time an integration over $\boldsymbol{\Omega}_c(\boldsymbol{r}, \tau)$ and $\boldsymbol{\Omega}_s(\boldsymbol{r}, \tau)$ with invariant integration measure normalized to unity, the partition function becomes

$$Z = \int \mathcal{D}\phi^*(\tau) \mathcal{D}\phi(\tau) \mathcal{D}\boldsymbol{\Omega}_c(\tau) \mathcal{D}\boldsymbol{\Omega}_s(\tau) \exp[-S(\phi^*, \phi, \boldsymbol{\Omega}_c, \boldsymbol{\Omega}_s)] \quad . \tag{51}$$

What has been done here is a change from a fixed reference frame in spin and particle–hole space to a reference frame varying in space and time. If one limits oneself to the spin rotation $SU(2)_s$, this amounts to using a quantization axis varying in space and time, as before. One can now proceed as in the previous chapter and for example integrate out the lower Hubbard band. In this way one can then solve the problem mentioned at the end of the last chapter.

The $SU(2)_c \times SU(2)_s$ symmetric formulation can also be used to obtain an effective Ginzburg–Landau type functional that contains both particle–hole and particle–particle type fields and recovers Hartree–Fock as its saddle point. For this purpose the Hamiltonian is written as

$$\begin{aligned} &H(\phi^*, \phi, \boldsymbol{\Omega}_c, \boldsymbol{\Omega}_s) = \\ &\quad -t \sum_{\langle \boldsymbol{r}\boldsymbol{r}' \rangle} \mathrm{tr}[R_{s\boldsymbol{r}} \Phi_{\boldsymbol{r}}^+ R_{c\boldsymbol{r}} R_{c\boldsymbol{r}'}^+ \Phi_{\boldsymbol{r}'} R_{s\boldsymbol{r}'}^+] + \frac{U}{16} \sum_{\boldsymbol{r}} \left\{ (\mathrm{tr}[\Phi_{\boldsymbol{r}}^+ \sigma_z \Phi_{\boldsymbol{r}}])^2 - (\mathrm{tr}[\sigma_z \Phi_{\boldsymbol{r}}^+ \Phi_{\boldsymbol{r}}])^2 \right\} \end{aligned} \tag{52}$$

The non-invariant way of writing the interaction part of $H$ is motivated by the fact that the saddle point equations for the resulting functional integral will reproduce the full Hartree-Fock equations.

The freedom gained by the introduction of the integrations over $\boldsymbol{\Omega}_c$ and $\boldsymbol{\Omega}_s$ will now allow me to restore the symmetries apparently lost in (52). First, I reintroduce the old variables $\Psi$ instead of $\Phi$. Thus

$$Z = \int \mathcal{D}\psi^*(\tau) \mathcal{D}\psi(\tau) \mathcal{D}\boldsymbol{\Omega}_c(\tau) \mathcal{D}\boldsymbol{\Omega}_s(\tau) \exp[-S(\psi^*, \psi, \boldsymbol{\Omega}_c, \boldsymbol{\Omega}_s)] \ , \tag{53}$$

with

$$S = \int_0^\beta d\tau \left[ \frac{1}{2} \sum_{\boldsymbol{r}} \mathrm{tr}[\Psi_{\boldsymbol{r}}^+ (\partial_\tau - \mu \sigma_z) \Psi_{\boldsymbol{r}}] + H(\psi^*, \psi, \boldsymbol{\Omega}_c, \boldsymbol{\Omega}_s) \right] \quad . \tag{54}$$

The Hamiltonian now is

$$\begin{aligned} &H(\psi^*, \psi, \boldsymbol{\Omega}_c, \boldsymbol{\Omega}_s) = \\ &\quad -t \sum_{\langle \boldsymbol{r}\boldsymbol{r}' \rangle} \mathrm{tr}[\Psi_{\boldsymbol{r}}^+ \Psi_{\boldsymbol{r}'}] + \frac{U}{16} \sum_{\boldsymbol{r}} \left\{ (\mathrm{tr}[\Psi_{\boldsymbol{r}}^+ \boldsymbol{\Omega}_c \cdot \boldsymbol{\sigma} \Psi_{\boldsymbol{r}}])^2 - (\mathrm{tr}[\boldsymbol{\Omega}_s \cdot \boldsymbol{\sigma}^T \Psi_{\boldsymbol{r}}^+ \Psi_{\boldsymbol{r}}])^2 \right\} \quad . \end{aligned} \tag{55}$$

The $SU(2)_c \times SU(2)_s$ symmetry is restored due to the integrations over $\mathbf{\Omega}_{c,s}$.

At first sight, eqs. (53), (54), and (55) look like a rather complicated way to rewrite the initial problem. The usefulness of the new representation will become apparent introducing a Hubbard-Stratonovich decomposition of the four-fermion interaction. One then obtains

$$Z = \int \mathcal{D}\Delta_s(\tau)\mathcal{D}\Delta_c(\tau)\mathcal{D}\mathbf{\Omega}_c(\tau)\mathcal{D}\mathbf{\Omega}_s(\tau) \exp[-S_{eff}(\mathbf{\Delta}_c, \mathbf{\Delta}_s)] \quad , \tag{56}$$

where $\mathbf{\Delta}_{c,s} = \Delta_{c,s}\mathbf{\Omega}_{c,s}$. Note that $\Delta_{c,s}$ is integrated over $(-\infty, \infty)$. The effective action for $\mathbf{\Delta}_{c,s}$ is (for $U > 0$)

$$\begin{aligned} &S_{eff}(\mathbf{\Delta}_c, \mathbf{\Delta}_s) = \\ &\quad \int_0^\beta d\tau \frac{1}{U} \sum_{\boldsymbol{r}} (\mathbf{\Delta}_{c\boldsymbol{r}}^2 + \mathbf{\Delta}_{s\boldsymbol{r}}^2) - \ln\left\{ \int \mathcal{D}\psi^*(\tau)\mathcal{D}\psi(\tau) \exp[-S(\psi^*, \psi, \mathbf{\Delta}_c, \mathbf{\Delta}_s)] \right\} \end{aligned} \quad , \tag{57}$$

with

$$\begin{aligned} &S(\psi^*, \psi, \mathbf{\Delta}_c, \mathbf{\Delta}_s) = \\ &\quad \int_0^\beta d\tau \left[ \frac{1}{2}\sum_{\boldsymbol{r}} \mathrm{tr}[\Psi_{\boldsymbol{r}}^+(\partial_\tau - \mu\sigma_z)\Psi_{\boldsymbol{r}}] - t \sum_{\langle \boldsymbol{r}\boldsymbol{r}'\rangle} \mathrm{tr}[\Psi_{\boldsymbol{r}}^+\Psi_{\boldsymbol{r}'}] \right. \\ &\quad \left. + \frac{1}{2}\sum_{\boldsymbol{r}} \left\{ i\mathbf{\Delta}_{c\boldsymbol{r}} \cdot \mathrm{tr}[\Psi_{\boldsymbol{r}}^+ \boldsymbol{\sigma} \Psi_{\boldsymbol{r}}] + \mathbf{\Delta}_{s\boldsymbol{r}} \cdot \mathrm{tr}[\boldsymbol{\sigma}^T \Psi_{\boldsymbol{r}}^+ \Psi_{\boldsymbol{r}}] \right\} \right] \quad . \end{aligned} \tag{58}$$

The saddle point equations now reproduce the full Hartree-Fock equations:

$$(\partial_\tau - \mu)\psi_{\boldsymbol{r}s} - t\sum_{\boldsymbol{r}'} \psi_{\boldsymbol{r}'s} + i\sum_{s'} (\mathbf{\Delta}_{c\boldsymbol{r}} \cdot \boldsymbol{\sigma})_{ss'}\chi_{\boldsymbol{r}s'} + \sum_{s'} (\mathbf{\Delta}_{s\boldsymbol{r}} \cdot \boldsymbol{\sigma})_{ss'}\psi_{\boldsymbol{r}s'} = 0 \quad , \tag{59}$$

$$\frac{2}{U}\mathbf{\Delta}_{c\boldsymbol{r}} = -\frac{i}{2}\langle \mathrm{tr}[\Psi_{\boldsymbol{r}}^+ \boldsymbol{\sigma} \Psi_{\boldsymbol{r}}] \rangle \ , \quad \frac{2}{U}\mathbf{\Delta}_{s\boldsymbol{r}} = -\frac{1}{2}\langle \mathrm{tr}[\boldsymbol{\sigma}^T \Psi_{\boldsymbol{r}}^+ \Psi_{\boldsymbol{r}}] \rangle \quad , \tag{60}$$

Here $\chi_{\boldsymbol{r}} = (\psi_{\boldsymbol{r}s}, s(-1)^{\boldsymbol{r}}\psi^*_{\boldsymbol{r},-s})$. Obviously (59) is the effective "Schrödinger equation" for electrons moving in the effective space- and time-dependent fields $\mathbf{\Delta}_{c,s}$, whereas eqs.(60) are the corresponding self-consistency conditions.

Though the Hartree–Fock equations (59) and (60) are rather standard, expanding around this saddle point gives rather interesting results. As far as the spin field $\mathbf{\Delta}_s$ is concerned, one finds a rather standard form. However, the transverse components of $\mathbf{\Delta}_c$ contain particle–parrticle excitations, and one therefore obtains a functional that allows one to study the reciprocal effects of particle–hole and particle–particle excitations on each other. This is potentially useful in studying possibilities of magnetically induced superconductivity or the occurence or not of magnetic instabilities in relatively dilute systems, where particle–particle like diagrams play an important role in renormalizing the effective interactions.

**Relation with the Slave–Fermion Formalism**

The rotation matrices (21) are elements of $SU(2)/U(1)$, and in particular have real diagonal elements. In the partition function (23) instead of integrating over this manifold, one can choose to integrate (with the properly normalized integration measure) over all $SU(2)$ matrices, i.e. one writes

$$R_{\boldsymbol{r}} = \begin{pmatrix} z_{1\boldsymbol{r}} & z^*_{2\boldsymbol{r}} \\ -z_{2\boldsymbol{r}} & z^*_{1\boldsymbol{r}} \end{pmatrix} \quad , \tag{61}$$

with the constraint $|z_{1\boldsymbol{r}}|^2 + |z_{2\boldsymbol{r}}|^2 = 1$. Transforming to new variables $b_{i\boldsymbol{r}}$ via

$$b_{i\boldsymbol{r}} = z_{i\boldsymbol{r}}(1 - \tfrac{1}{2}\phi^*_{\boldsymbol{r}}\phi_{\boldsymbol{r}}) \tag{62}$$

the constraint becomes

$$|b_{1\boldsymbol{r}}|^2 + |b_{2\boldsymbol{r}}|^2 + \phi^*_{\boldsymbol{r}}\phi_{\boldsymbol{r}} = 1 \tag{63}$$

This is nothing but the constraint familiar from the slave–fermion formulation of the strongly correlated fermion problem: on each site there is either a spin (represented here by a boson of spin ↑ ($b_1$) or ↓ ($b_2$)) or a hole, represented here by a spinless fermion. Performing the same transformation on the action, one finds

$$\begin{aligned} S \;=\; \int_0^\beta d\tau \Bigg\{ & \sum_{\boldsymbol{r}} (b^*_{s\boldsymbol{r}}\partial_\tau b_{s\boldsymbol{r}} + \phi^*_{\boldsymbol{r}}(\partial_\tau - \mu)\phi_{\boldsymbol{r}}) - t \sum_{\langle \boldsymbol{r}\boldsymbol{r}'\rangle} (b^*_{s\boldsymbol{r}} b_{s\boldsymbol{r}'}\phi^*_{\boldsymbol{r}'}\phi_{\boldsymbol{r}} + c.c.) \\ & + \frac{J}{4} \sum_{\langle \boldsymbol{r}\boldsymbol{r}'\rangle} (b^*_{s\boldsymbol{r}}\boldsymbol{\sigma}_{ss'} b_{s'\boldsymbol{r}}) \cdot (b^*_{t\boldsymbol{r}'}\boldsymbol{\sigma}_{tt'} b_{t'\boldsymbol{r}'}) + \ldots \Bigg\} \quad , \end{aligned} \tag{64}$$

where the omitted terms are three–site terms analogous to those in eq.(44). This is nothing but the action of the $t-J$ model in the slave–fermion representation.[24] Note that the factors $(1 - \phi^*_{\boldsymbol{r}}\phi_{\boldsymbol{r}})$ and $(2 - \phi^*_{\boldsymbol{r}}\phi_{\boldsymbol{r}} - \phi^*_{\boldsymbol{r}'}\phi_{\boldsymbol{r}'})$ which appeared in (41) and (44) have now disappeared. The same reasoning can be followed in the more general $SU(2) \times SU(2)$ formulation, and one then obtains other types of slave–particle boson or fermion) representations.